\documentclass[traditabstract]{aa}

\usepackage{graphicx}
\usepackage{txfonts}
\usepackage{color}
\usepackage[draft]{hyperref}
\usepackage{natbib}
\bibpunct{(}{)}{;}{a}{}{,} 
\usepackage[english]{babel}
\newcommand{\beqa}{\begin{eqnarray}} 

\newcommand{\eeqa}{\end{eqnarray}}

\newcommand{\bsub}{\begin{subequations}}
\newcommand{\esub}{\end{subequations}}
\newcommand{\beal}{\begin{align}}
\newcommand{\ealn}{\end{align}}

\newcommand{\Nif}{$\rm ^{56}Ni$\,}

\def\gsim{\mathrel{\rlap{\lower 4pt \hbox{\hskip 1pt $\sim$}}\raise 1pt \hbox {$>$}}}
\def\lsim{\mathrel{\rlap{\lower 4pt \hbox{\hskip 1pt $\sim$}}\raise 1pt \hbox {$<$}}}


\begin{document}

\title{Nebular spectroscopy of SN 2014J: Detection of stable nickel in near infrared spectra}

\titlerunning{NIR nebular spectra SN 2014J}
\authorrunning{Dhawan et al.}
\author{\textbf{S. Dhawan\inst{1,2}
        \and A. Fl\"ors\inst{2,3,4}
        \and B. Leibundgut\inst{2,5}
        \and K. Maguire\inst{6}
        \and W. Kerzendorf\inst{2}
        \and S. Taubenberger\inst{2,3}
        \and M. Van~Kerkwijk\inst{7}
        \and J. Spyromilio\inst{2}
    }}

\institute{
Oskar Klein Centre, Department of Physics, Stockholm University,
SE 106 91 Stockholm, Sweden \\
\email{suhail.dhawan@fysik.su.se}
\and European Southern Observatory, Karl-Schwarzschild-Strasse 2,
D-85748 Garching bei M\"unchen, Germany 
\and Max-Planck-Institut f\"ur Astrophysik, Karl-Schwarzschild-Stra\ss e 1,
D-85748 Garching bei M\"unchen, Germany
\and Physik-Department, Technische Universit\"at M\"unchen, James-Franck-Stra\ss e 1, D-85748 Garching bei M\"unchen, Germany 
\and  Excellence Cluster Universe, Technische Universit\"at M\"unchen,
Boltzmannstra\ss e 2, D-85748, Garching, Germany
\and Astrophysics Research Centre, School of Mathematics and Physics, Queen's University Belfast, Belfast BT7 1NN, UK
\and Department of Astronomy and Astrophysics, University of Toronto, 50 St. George Street, Toronto, ON M5S 3H4, Canada
} 

\date{Received; accepted }

\offprints{S.~Dhawan}

 \abstract{We present near infrared (NIR) spectroscopy of the nearby supernova 2014J obtained $\sim$450\,d after explosion. We detect the [\ion{Ni}{II}] 1.939\,$\mu$m line in the spectra indicating the presence of stable \element[][58]{Ni} in the ejecta. The stable nickel is not centrally concentrated but rather distributed as the iron. The spectra are dominated by forbidden [\ion{Fe}{II}] and [\ion{Co}{II}] lines. We use lines, in the NIR spectra, arising from the same upper energy levels to place constraints on the extinction from host galaxy dust. We find that that our data are in agreement with the high $A_V$ and low $R_V$  found in  earlier studies from data near maximum light.  Using a  \Nif\, mass prior from near maximum light $\gamma$-ray observations, we find  $\sim$0.05\,M$_\odot$ of stable nickel to be present in the ejecta. We find that the iron group features are redshifted from the host galaxy rest frame by $\sim$600\,km\,s$^{-1}$. 
}

\maketitle 
\section{Introduction}
Type Ia supernovae (SNe\,Ia) have long been identified as thermonuclear explosions of  C/O white dwarfs \citep[WD;][]{1960ApJ...132..565H}. Following a simple calibration \citep{1993ApJ...413L.105P} they are excellent distance indicators used extensively in cosmology \citep{1998AJ....116.1009R,1999ApJ...517..565P}. There remain several open questions regarding the physics of SNe\,Ia, e.g. progenitor channel, mass of the progenitor, explosion mechanism, (see \cite{2000ARA&A..38..191H, 2001ARA&A..39...67L,2013FrPhy...8..116H,2014ARA&A..52..107M,2018arXiv180203125L} for reviews). Many attempts to address these issues have concentrated on photometric and spectroscopic observations of SNe\,Ia near maximum light (i.e. the photospheric phase). However, at late times, as the ejecta thin and the core of the ejecta is revealed, additional diagnostics of the explosion mechanism become accessible. The elegance of SNe\,Ia lies in the fact that both the energy of the explosion and the electromagnetic display are due to the burning of the core to \element[][56]{Ni} and its subsequent decay.  Direct or indirect measurements of the mass and topology of \element[][56]{Ni} provide some of the best probes of the explosion physics. The decay of \element[][56]{Ni} to \element[][56]{Co} and on to stable \element[][56]{Fe} proceeds through the emission of $\gamma$-rays and positrons which power the electromagnetic display.

Nebular phase spectroscopy enables a direct view into the core of the ejecta and provides key insights into the progenitor and explosion properties of SNe\,Ia, complementary to the early time observations which contain information about the outer layers of the ejecta. The evolution of the cobalt to iron line ratios in nebular spectra have been used to demonstrate radioactive decay as the mechanism powering SNe\,Ia \citep{1994ApJ...426L..89K} and a combination of optical and NIR spectra have provided constraints on the iron mass in the ejecta \citep{1992MNRAS.258P..53S,2004A&A...426..547S}. Moreover, correlations of nebular-phase line velocities with photospheric-phase velocity gradients have indicated asymmetries in the explosion mechanism \citep{2010Natur.466...82M,2010ApJ...708.1703M,2018MNRAS.tmp..796M}. Most nebular-phase studies of SNe\,Ia \citep[e.g.][]{2013MNRAS.430.1030S,2017MNRAS.472.3437G} have concentrated on optical wavelengths. Since the iron and cobalt features in the NIR are relatively unblended compared to the optical, it is an interesting wavelength region to study the line profiles. Some studies have used the [\ion{Fe}{II}]\,1.644\,$\mu$m feature to probe the kinematic distribution of the radioactive ejecta \citep{2004ApJ...617.1258H,2006ApJ...652L.101M} and have also tried to  constrain  the central density and magnetic field of the progenitor WD \citep{2014ApJ...795...84P,2015ApJ...806..107D}. \citet{2007ApJ...661..995G} have used mid-IR nebular spectra to explore the explosion mechanism and electron capture elements.  

The presence of large amounts of stable isotopes of nickel (e.g. \element[][58]{Ni} and \element[][60]{Ni}) has been suggested as an indicator of burning at high central densities and therefore would favour higher progenitor masses, contributing to evidence for a Chandrasekhar-mass progenitor scenario for some supernovae \citep[][]{2004ApJ...617.1258H}.

SN\,2014J in M82 is one of the nearest SN\,Ia in decades and has been extensively studied \citep[e.g.][]{2014ApJ...790....3K,2014Sci...345.1162D,2014MNRAS.443.2887F,2014MNRAS.445.4427A,2015ApJ...804...66V,2016MNRAS.457..525G,2016MNRAS.460.1614V}. It is heavily reddened by host galaxy dust, but shows spectral features of a normal SN\,Ia \citep{2014ApJ...788L..21A}. The extinction law and properties of the dust along the line of sight have been extensively discussed by a number of authors (\citet{2014ApJ...795L...4K}, \citet{2014MNRAS.443.2887F}, \citet{2014ApJ...784L..12G}). Imaging of the supernova reveals the presence of clear echoes \citep{2017ApJ...834...60Y}.  
Estimates of the $^{56}$Ni\, mass using the timing of the NIR second maximum \citep{2016A&A...588A..84D} and $\gamma$-ray observations  \citep{2015ApJ...812...62C} infer masses of 0.64\,$\pm$\,0.13\,$M_{\odot}$ and 0.62\,$\pm$\,0.13 $M_{\odot}$, respectively, consistent with the estimate for a normal SN\,Ia \citep{2006A&A...450..241S,2014MNRAS.440.1498S}. Optical spectra of SN\,2014J in the nebular phase are remarkably similar to the `normal' SNe\,2011fe and 2012cg \citep{2015MNRAS.453.3300A}. 

In this work we present the first detection of a spectral line at 1.939\,$\mu$m in an SN\,Ia, a forbidden [\ion{Ni}{II}] transition ($^4$F$_{9/2}$--$^2$F$_{7/2}$), indicating the presence of stable nickel isotopes \citep[e.g., see][]{2018MNRAS.474.3187W}. We use the ratios of the [\ion{Fe}{II}] lines to measure the parameters describing the host galaxy extinction and evaluate the line shifts and profiles for the spectra. We present the data in Section~\ref{sec-obs}, analyse them in Section~\ref{sec-ana},  discuss our results in Section \ref{sec-dis} and conclude in Section \ref{sec-conc}. 

\begin{table*}
\begin{minipage}{160mm}
\begin{center}
\caption{Observing log of spectra obtained with GNIRS on the Gemini-North telescope.}
\begin{tabular}{llccrrr}
MJD & UT Date & Phase & Wavelength coverage ($\mu$m) & Exposure Time (s)\\
\hline
\\
57096.085 & Mar 15, 2015 & +408 &0.825 - 2.5 & 2400\\
57137.864 & Apr 26, 2015 & +450 &0.825 - 2.5  &2400\\
57166.263 & May 24, 2015 & +478 &0.825 - 2.5	& 2400\\
\hline
\end{tabular}
\end{center}
\end{minipage}
\label{tab:log}
\end{table*}

\section{Observations and data reduction}
\label{sec-obs}

SN\,2014J was discovered in M82 at the University of London Observatory on Jan 21st 2014 \citep{2014CBET.3792....1F}. We adopt MJD56671.7 as the epoch of explosion \citep{2014ApJ...784L..12G}. We present NIR spectra of SN\,2014J, obtained using GNIRS on Gemini-North. The dates and phases are shown in Table\,\ref{tab:log}. The observations were made using the Gemini fast turnaround program \citep{2014SPIE.9149E..10M} under proposal GN-2015A-FT-3. The spectra were obtained in cross-dispersed (XD) mode with a central wavelength  of 1.65\,$\mu$m and have a wavelength coverage from 0.825 to 2.5\,$\mu$m. 

The spectra were reduced using the standard Gemini IRAF\footnote{IRAF is distributed by the National Optical Astronomy Observatories, which are operated by the Association of Universities for Research in Astronomy, Inc., under cooperative agreement with the National Science Foundation} package. The final spectra were extracted using the IRAF task \emph{apall} which extracts one-dimensional sums across the apertures. For the first epoch (March) we used the A0 star HIP\,32549 from the Gemini archive (not observed on the same night) for telluric correction corrections but we are unable to colour correct our data since no suitable standard on the same night was observed. The March spectrum is therefore only used for measurements of line shifts and profiles but not for line ratios. For the second (April) and third (May) epochs, we use the A7 star HIP\,50685 obtained on the same nights as the observations of the supernova for both telluric and spectrophotometric calibration. Given that the features we are studying lie in regions of variable atmospheric transmission, we have checked our standard star observations against atmospheric transmission models of the atmosphere. We find that the transmission of the standard star is compatible with the Mauna Kea atmospheric models from the Gemini web site\footnote{http://www.gemini.edu/sciops/telescopes-and-sites/observing-condition-constraints/ir-transmission-spectra} for 50 mm of water vapour and an airmass of 2.0. The individual features of the atmosphere are reproduced well by the spectrum at the appropriate instrumental resolution.

\begin{figure*}
\centering
\includegraphics[angle=-0, width=.98\textwidth]{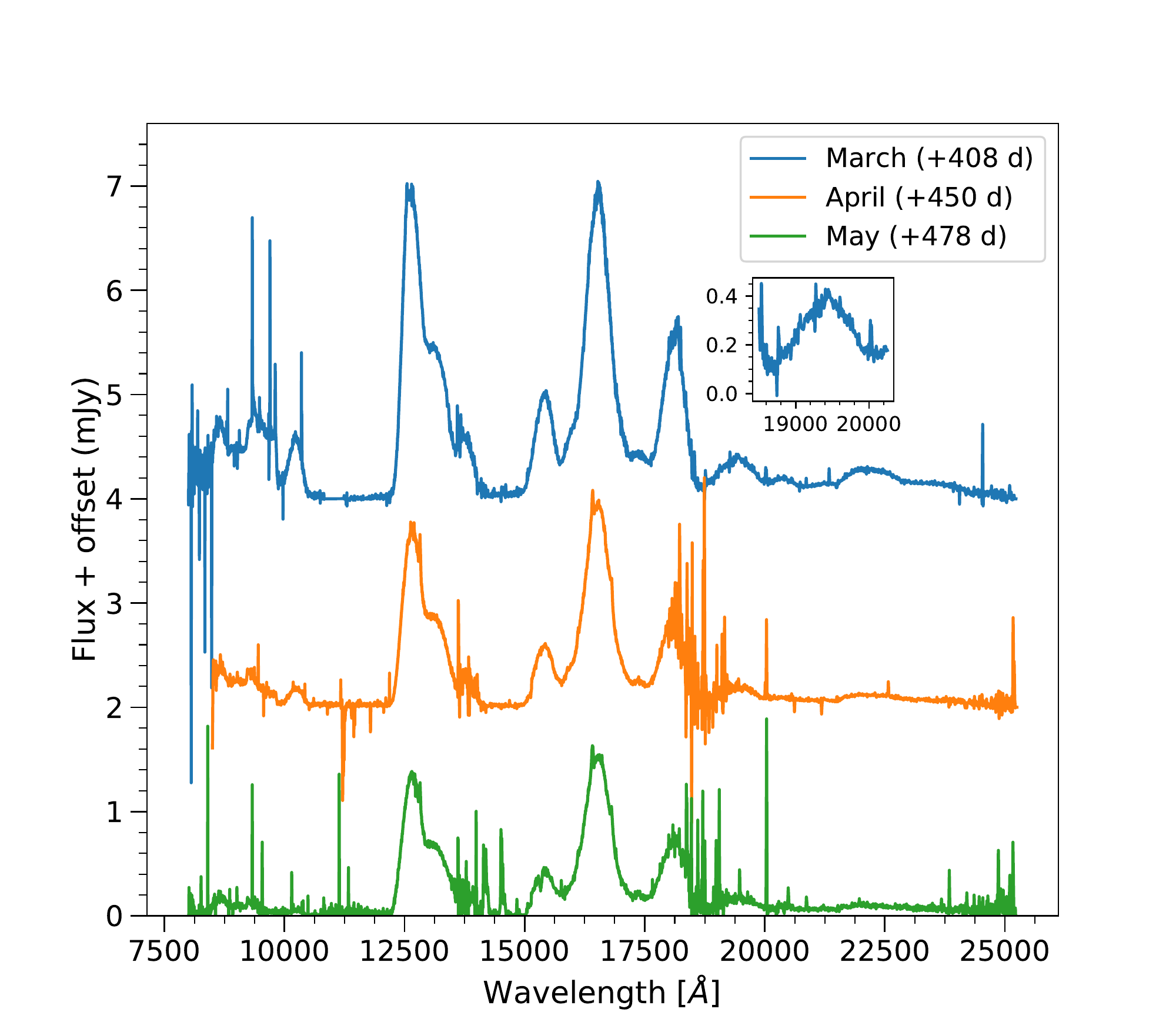}
\caption{NIR spectra of SN\,2014J (for the three nights it was observed). The spectra are plotted in mJy for visibility and each spectrum is incrementally displaced vertically by 2\,mJy. The March data are at top and the May data at the bottom. The regions of poor atmospheric transmission around 1.4 and 1.9\,$\mu$m are clearly identifiable due to the increased noise. The inset shows the region between 1.85 and 2.03 $\mu$m (the expected region for the [Ni II] feature) in the March spectrum.}
\label{fig:spectra}
\end{figure*}

No discernible continuum is present in the $J$ and $H$ bands in any of our data. In the $K$ band we cannot exclude the presence of an underlying continuum although this may have an instrumental contibution as well as an  astronomical one. The spectra are shown in Figure\,\ref{fig:spectra}.

\section{Analysis}\label{sec-ana}

\subsection{Nickel detection}

\begin{figure}
\includegraphics[width=.47\textwidth]{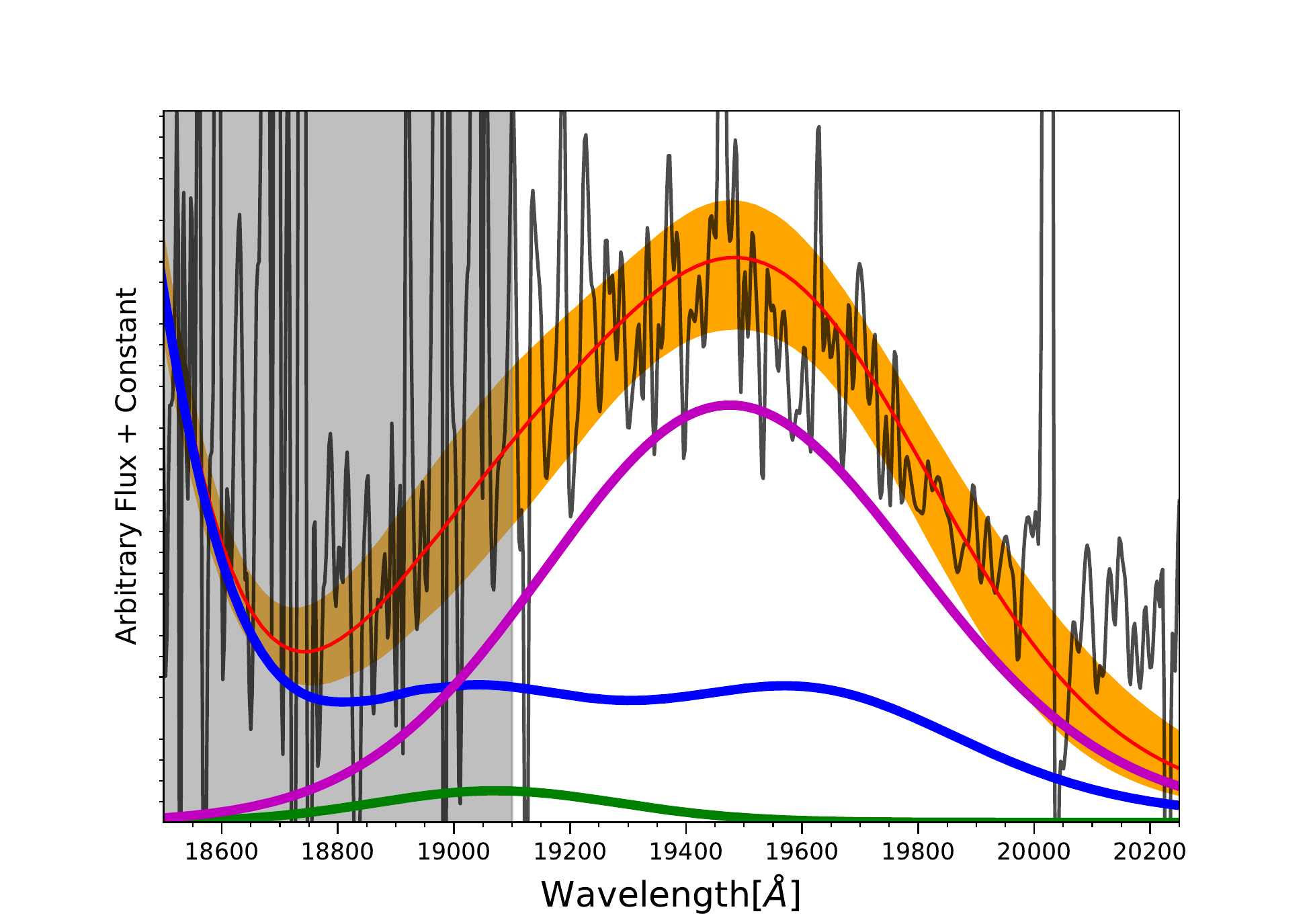}
\caption{The May spectrum and a model using an [\ion{Ni}{II}] NLTE atom (magenta) with a line width of 11\,000\,km\,s$^{-1}$ redshifted by 800\,km\,s$^{-1}$ from the laboratory rest frame. In the  1.8--2\,$\mu$m region weak [\ion{Fe}{II}] lines (blue) and [\ion{Co}{II}] lines (green) are also present. Shortwards of the central wavelength of this line the transmission is low and the noise in the spectrum increases (grey shaded region). This region has not been used in the fitting algorithm.}
\label{fig:model_ni2}
\end{figure}

In our data (see Figure~\ref{fig:spectra}), we note the presence of a weak line at a wavelength coincident with the central wavelength of a [\ion{Ni}{II}] transition. In Figure~\ref{fig:model_ni2} we fit the feature and present the first clear detection of the $^4$F$_{9/2}$--$^2$F$_{7/2}$ [\ion{Ni}{II}] 1.939 $\mu$m line in the spectra of a Type~Ia supernova. Since the e-folding timescale of radioactive nickel is 8.8\,d \citep{1994ApJS...92..527N} and the spectra are taken at $\sim$ 450\,d, this would indicate that the detection of the [\ion{Ni}{II}] line indicates the presence of a stable nickel isotope. 

In our spectra there is no evidence for the  $^2$D$_{5/2}$--$^4$F$_{7/2}$ line of [\ion{Ni}{II}] at 1.0718\,$\mu$m, which has an A value two orders of magnitude lower than the 1.939\,$\mu$m line. Our modeling of the atmospheric transmission and the standard star spectra suggest that, if the line were strong, we should observe the center of the line and the blue wing. Given the extinction and our assumed excitation conditions (see Section~\ref{fit}) we cannot use the absence of a strong line at 1.0718\,$\mu$m to challenge the identification of the 1.939\,$\mu$m line. 												


\subsection{Fitting the spectrum}\label{fit}

The NIR spectrum of SN\,2014J is dominated by forbidden lines of singly ionised iron and singly ionised cobalt. In particular the prominent features at the long end of the $J$ window (around 1.26 $\mu$m) and most of the $H$-band emission arise from multiplets a$^6$D--a$^4$D and a$^4$F--a$^4$D of [\ion{Fe}{II}], respectively, while the feature around 0.86\,$\mu$m has contributions from the [\ion{Fe}{II}] a$^4$F--a$^4$P multiplet and from [\ion{Co}{II}] a$^3$F--b$^3$F. Additional contributions from  [\ion{Co}{II}] a$^5$F--b$^3$F are present in the $H$ band (at 1.547\,$\mu$m). The line identification is provided in Table \ref{tab:lines}. 

\begin{table}
\begin{minipage}{70mm}
\centering
\tiny
\caption{Dominant line identifications for SN\,2014J }
\scalebox{.9}{\begin{tabular}{|l|l|c|c|c|}
\hline

$\lambda$ & Species & transition & flux 450\,d & flux 478\,d   \\
$\mu$m & & & erg/s/cm$^2$ & erg/s/cm$^2$ \\
\hline
&&&&\\
0.8617	& [\ion{Fe}{II}] & a$^4$F$_{9/2}$--a$^4$P$_{5/2}$ & (4.26$\pm$0.74)$\times$10$^{-14}$ & (2.45$\pm$0.25)$\times$10$^{-14}$   \\
1.019  & [\ion{Co}{II}] & a$^3$F$_4$--b$^3$F$_4$  & (1.40$\pm$0.12)$\times$10$^{-14}$ & (4.24$\pm$0.35)$\times$10$^{-15}$ \\
1.257  & [\ion{Fe}{II}] & a$^6$D$_{9/2}$--a$^4$D$_{7/2}$ & (1.20 $\pm$0.07)$\times$10$^{-13}$ & (9.89$\pm$0.16)$\times$10$^{-14}$ \\
1.547  & [\ion{Co}{II}] & a$^5$F$_5$--b$^3$F$_4$  & (9.35$\pm$0.32)$\times$10$^{-15}$ & (2.83$\pm$0.23)$\times$10$^{-15}$ \\
1.644  & [\ion{Fe}{II}] & a$^4$F$_{9/2}$--a$^4$D$_{7/2}$  & (9.62$\pm$0.21)$\times$10$^{-14}$ & (7.91$\pm$0.13)$\times$10$^{-14}$ \\
1.939  & [\ion{Ni}{II}] & a$^2$F$_{7/2}$--a$^4$F$_{9/2}$ & (1.03$\pm$0.11)$\times$10$^{-14}$ & (9.68$\pm$0.87)$\times$10$^{-15}$ \\
\hline
\end{tabular}}
\label{tab:lines}
\end{minipage}
\end{table}

Fitting the $J$ and $H$ band spectra is relatively straightforward. The $K$ band is very faint but the transitions in the $K$ band are not expected to be strong. The model we use (Fl\"ors et al. in preparation) uses NLTE excitation of singly and doubly ionised iron group elements. For the purposes of fitting only the NIR data, the only contributing features are from the singly ionised species and a one zone model, convolved with a Gaussian line profile, suffices to provide an excellent fit. The atomic data for our NLTE models are from \citet{2016MNRAS.456.1974S}, \citet{1988A&A...193..327N}, \citet{1980A&A....89..308N}, \citet{2010A&A...513A..55C}. 
The free parameters are the electron density and temperature of the gas, the ratio of iron to cobalt to nickel, the line width, the offset from the systemic velocity of the supernova that the singly ionised transitions of the iron group elements may exhibit and the extinction. We explore the parameter space with the nested-sampling algorithm Nestle (https://github.com/kbarbary/nestle) and a $\chi^2$ likelihood. Uniform priors are used for all parameters except the electron density. For this we assume a lower bound of 10$^5$\,cm$^{-3}$. As has also been observed for other SN\,Ia at this epoch, our spectrum shows no evidence for neutral iron lines. The strong lines of the [\ion{Fe}{I}] a$^5$D--a$^5$F multiplet at 1.4\,$\mu$m are in a poor transmission region but still remain undetected. The lowest-excitation and therefore presumably strongest [\ion{Fe}{I}] line in the 2\,$\mu$m region at 1.9804\,$\mu$m from multiplet a$^5$F--a$^3$F is also not seen.  The infrared spectrum provides some evidence in the $K$ band for doubly ionised iron but it is well known from combined optical and infrared spectra \citep[e.g.][]{1992MNRAS.258P..53S} that strong transitions of [\ion{Fe}{III}] and [\ion{Co}{III}] are present in the 4000 to 6000\,\AA\ region.  As discussed in the introduction, from earlier work \citep{2015ApJ...812...62C, 2016A&A...588A..84D} it has been determined that SN\,2014J made  $\approx$0.6\,M$_\odot$ of $^{56}$Ni. A simple distribution of this mass of Nickel and its daughter elements in singly ionised state in a volume expanding for the age of the supernova at 8000\,km\,s$^{-1}$ sets the electron density to be at least 10$^5$\,cm$^{-3}$. 

\begin{figure*}
\includegraphics[width=.98\textwidth]{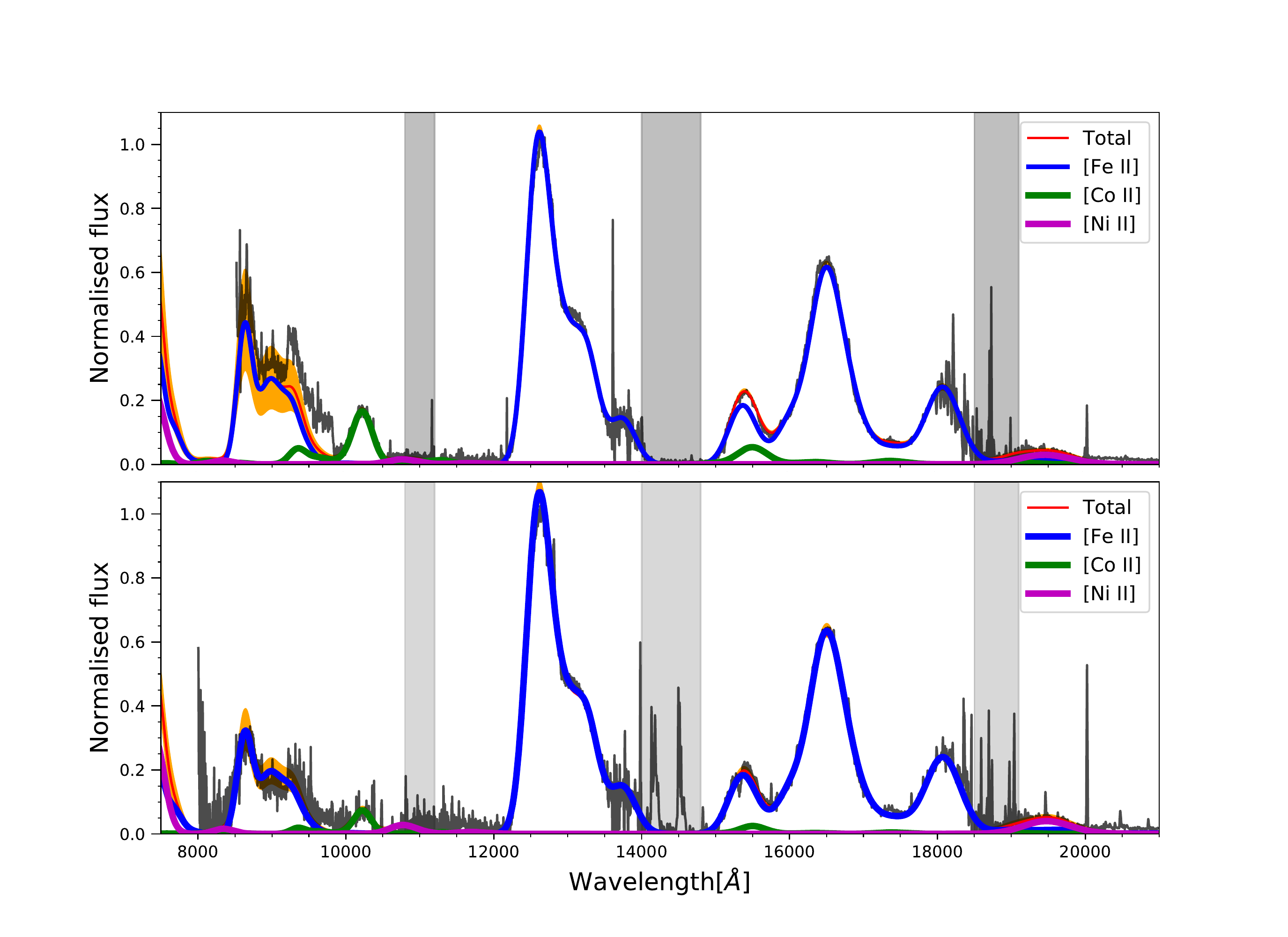}
\caption{Fits to the observations using a NLTE one zone emission code. (Top): April and (Bottom): May spectra. The red lines indicate the mean model. The orange shaded band shows the 95\% credibility region. Ion contributions are shown in blue ([\ion{Fe}{II}]), green ([\ion{Co}{II}]) and magenta ([\ion{Ni}{II}]). The atmospheric absorption bands shortwards of 1.12, 1.48 and 1.915\,$\mu$m are shaded grey. The resulting values for the temperature and electron density for April spectrum are 3700 $\pm$ 400 K and 2.18 ($\pm$ 0.56) $\cdot$ 10$^5$ cm$^{-3}$ and 3300 $\pm$ 200 K and 1.69 ($\pm$ 0.59) $\cdot$ 10$^5$ cm$^{-3}$}
\label{fig:fits}
\end{figure*}

The fits are shown in Figure~\ref{fig:fits}. Additional constraints arise from a few higher excitation lines in the 8600\,\AA\, region but these are extremely sensitive to the choice of atomic data. For the purposes of this work we note that we can fit these lines and that the fits are consistent with our derived properties but we do not draw conclusions based on this aspect of our data.

\subsection{Extinction}
\label{ssec-ext}

\begin{figure}
\centering
\includegraphics[angle=-0, width=9cm]{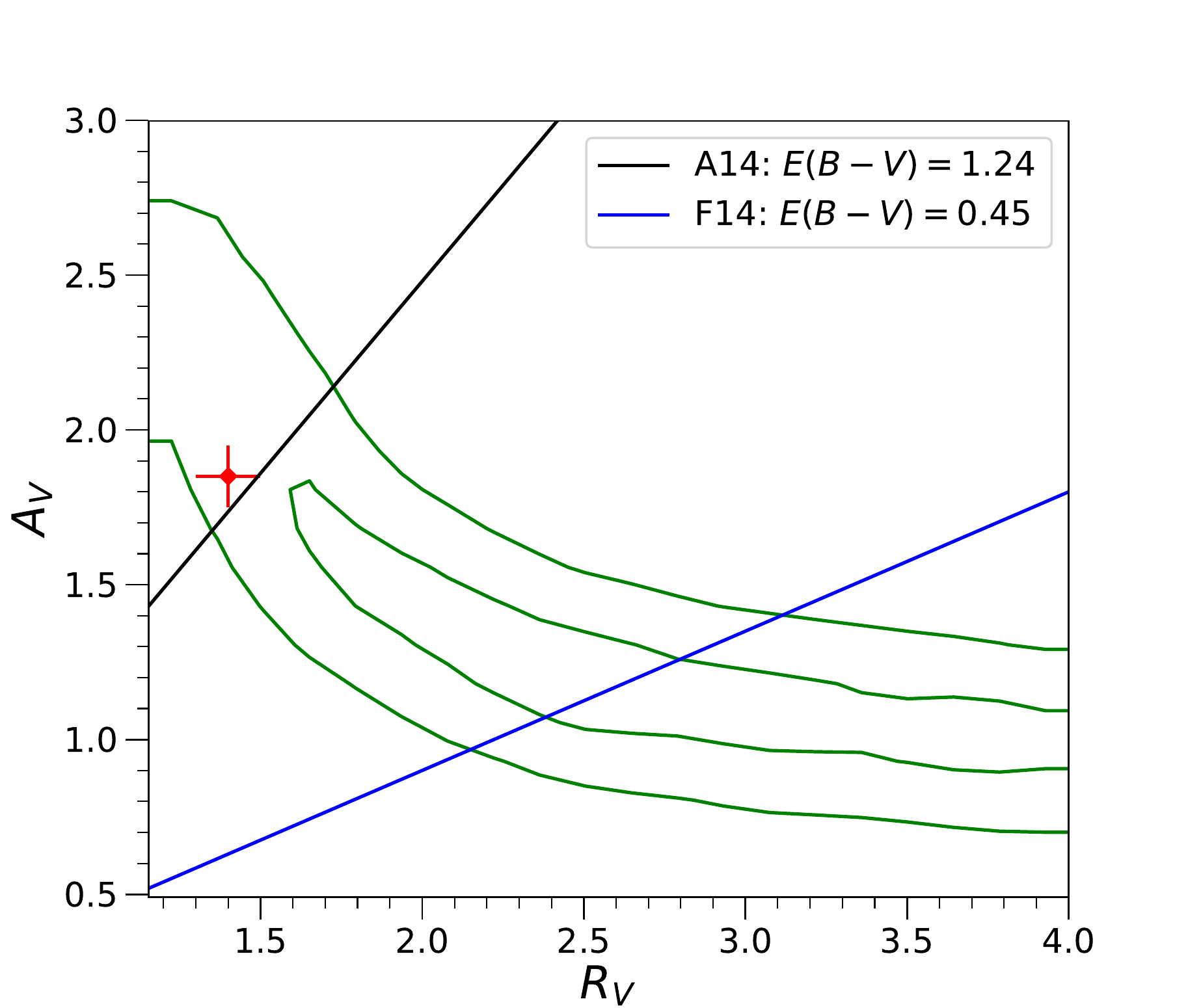}
\caption{The distribution for $R_V$ and $A_V$ values by fitting the [\ion{Fe}{II}] lines (the contours shown here are for the April spectrum, the results for the May spectrum are consistent with the contours displayed here). The values measured by \citet{2014ApJ...788L..21A} are marked in red and agree with our April spectrum. The contours mark the 1 and 2 $\sigma$ credible regions of the MCMC samples. The lines show a constant $E(B-V)$ of 1.37 as derived by \citet{2014ApJ...788L..21A} and a constant $E(B-V)$ of 0.45 as derived by \citet{2014MNRAS.443.2887F}.}
\label{fig:rv}
\end{figure}

The two strongest lines in the spectrum are the [\ion{Fe}{II}] 1.257\,$\mu$m\,  a$^6$D$_{9/2}$--a$^4$D$_{7/2}$ and the 1.644\,$\mu$m\, [\ion{Fe}{II}] a$^4$F$_{9/2}$--a$^4$D$_{7/2}$ which arise from the same upper level. In the absence of additional contributions to the features at these wavelengths and optical depth effects the observed line ratio and can be used to determine the extinction between the $J$ and $H$ bands as it depends solely on the Einstein A-values these transitions. Additional constraints, independent of the excitation conditions come from the [\ion{Co}{II}] lines at 1.547\,$\mu$m (a$^5$F$_4$--b$^3$F$_4$) and 1.0191\,$\mu$m (a$^3$F$_4$--b$^3$F$_4$) which also come from the same upper level. From our model fits of these features we can ensure that blending is taken consistently into account. We constrain the \citet{1989ApJ...345..245C} prescription for the extinction  in the range shown in  Figure\,\ref{fig:rv}. The NIR data are compatible with both high and low values of $R_V$. The degeneracy between $A_V$ and $R_V$ arises from the short wavelength lever arm between the $J$ and $H$ windows.

\subsection{Line shifts and widths}
As noted by \cite{2010ApJ...708.1703M} some type\,Ia supernovae exhibit a shift of the line centres of the singly ionised iron group lines relative to the systemic velocity of the supernova and also with respect to the doubly ionised features. In the absence of doubly ionised features in our data we cannot identify a differential velocity shift. However, the lines in our data exhibit a $\sim$800\,km\,s$^{-1}$ redshift which is $\sim$600\,km\,s$^{-1}$ in excess of the M82 recession velocity. The [\ion{Ni}{II}], [\ion{Co}{II}] and [\ion{Fe}{II}] lines exhibit the same shifts within the errors, although the constraint on the [\ion{Ni}{II}] is weak and degenerate with the width.

The profiles of the iron and cobalt lines are well fit by a simple Gaussian line shape for each individual component of the multiplets. We find that a Full Width Half Maximum of 8\,600$\pm$150\,km\,s$^{-1}$ fits the [\ion{Fe}{II}] and [\ion{Co}{II}] data. 

The [\ion{Ni}{II}]  line width  is $\sim$11\,000\,km\,s$^{-1}$, somewhat higher than that needed to fit the iron and cobalt lines. It is thus clear that the stable iron group elements are not at the lowest velocities, unlike the predictions from 1-D $M_{ch}$ models. However, we find that the model predictions from 3-D delayed detonation explosions of \citet{2013MNRAS.429.1156S} which predict stable isotopes at intermediate velocities are consistent with our observations. The somewhat higher velocity of the Nickel may be an artifact of our continuum placement which has been conservatively assumed to be non-existent. A small residual continuum, possibly from incomplete background subtraction, would result in the [\ion{Ni}{II}] velocity being consistent with the other lines (see Section~\ref{fit}). 

\section{Discussion}
\label{sec-dis}

\citet{2018MNRAS.474.3187W} argued that the [\ion{Ni}{II}] 1.939 $\mu$m line is relatively unblended compared to the strong [\ion{Ni}{II}] feature at 7378\,\AA\, in the optical and therefore a more suitable test for the presence of large amounts of stable Nickel. \citet{2018MNRAS.474.3187W} used models of \citet{2013MNRAS.429.2127B} and \citet{2017MNRAS.470..157B} to model the optical and NIR spectra in the nebular regime. 

That study (see \citet{2018MNRAS.474.3187W} figure 13) shows a dramatic variation in the strength of the 1.939\,$\mu$m line for their models, however, we note that their synthetic spectra are at epochs $\sim$ 200\,d before the spectra presented here. The mass of stable nickel, however, in the models of \citet{2018MNRAS.474.3187W} only varies by a factor of 3 between the different models used ($\sim$0.011 to 0.03\,M$_\odot$). Compared to their yields, the $M_{ch}$ models of \citet{2013MNRAS.429.1156S} are slightly higher in the range between 0.03 and 0.07 $M_{\odot}$. The appearance of the spectrum is particularly sensitive to the ionisation conditions in the models and therefore the derivation of masses from lines arising from a single ionisation stage, such as we have attempted here, is challenging. Our confirmation of the presence of the 1.939\,$\mu$m line, which arises from the same upper level as the 7378\,\AA\ line confirms the identification of [\ion{Ni}{II}] by \citet{2018MNRAS.tmp..796M}.

\subsection{Ionisation and masses of Fe$^+$ and Ni$^+$}

From our fitting we determine an un-blended flux for 1.644\,$\mu$m line and infer the mass of the emitting iron. For a distance to M82 of 3.5\,Mpc  \citep{2006Ap.....49....3K} and our measured flux for the 1.644\,$\mu$m line we determine a mass of $\sim$0.18$\,M_{\odot}$ for Fe$^+$ at the epoch of the SN spectrum. We note that without the high density prior from the \Nif\, mass, our spectra allow for low densities and high excitation conditions  which can give Fe$^+$ yields as low as 0.01 M$_\odot$ (the masses reported here are the weighted mean for the +450 and +478\,d spectra).

We note the strength of the 1.533\,$\mu$m a$^4$F$_{9/2}$--a$^4$D$_{5/2}$ line and the presence  of a blue shoulder at the 1.644\,$\mu$m feature (at 1.6\,$\mu$m) due to the a$^4$F$_{7/2}$--a$^4$D$_{3/2}$ [\ion{Fe}{II}] line at 1.599\,$\mu$m. The ratios of these lines to the 1.644\,$\mu$m line are sensitive to the electron density, and the 1.54\,$\mu$m feature as well as the shoulder drop for electron densities below 10$^5$\,cm$^{-3}$ (see also \citet{1988A&A...193..327N}). Here again we have avoided placing a disproportionate weight on a particular feature in our likelihood function. We only note that this presents corroborating evidence for high electron densities, consistent with the inference on the density from the \Nif\, mass prior, hence increasing our confidence in the Fe$^{+}$ mass determination. 

Using the derived mass of Fe$^{+}$, the total iron mass prior and assuming the only the singly and doubly ionised species are present, we get an ionisation fraction of $\approx$1.7. This value is consistent with calculations from theory ($\sim$ 1.6) in the literature \citet[e.g.][]{1980PhDT.........1A}.


We can now proceed to derive a value of 0.016$\pm$0.005\,M$_\odot$ for the Ni$^+$ mass based on the emissivity of the 1.939\,$\mu$m line and the other derived parameters from our fits. Similarly, assuming that all the Ni exists in singly and doubly ionised state and that ionisation fractions for nickel as for iron \citep[see][for caveats]{2018MNRAS.474.3187W} then we estimate that approximately 0.053$\pm$0.018\,M$_\odot$ of stable nickel is present in SN\,2014J. While this is higher than the predictions for the different models, it is within 3 $\sigma$ of the predicted range of estimates. 

We observed a complex of weak [Fe III] lines in the $K$-band. Using the Fe$^{++}$ mass derived above (0.42 M$_\odot$), the distance to M82 (3.5 Mpc) and the observed flux in the region between 21000 and 24000 \AA\, ($0.75\pm0.2\times10^{-14}$ ergs$^{-1}$cm$^{-2}$) we derive an emissivity of $\sim 2 \cdot 10^{-18}$ erg s$^{-1}$atom$^{-1}$. This is consistent with the derived line emissivity from the NLTE calculations with a temperature of 4000 K and N$_e$ of 10$^5$ cm$^{-3}$. However, these estimates are highly sensitive to the temperature and N$_e$ values, placement of the continuum and hence, only offer a consistency check. 

\subsection{Extinction by host galaxy dust}

Measurements from maximum light photometry would point to a high $E(B-V)$ \citep[1.37 mag, see ][]{2014ApJ...788L..21A}. The authors find a non-standard reddening law describing the colour excesses at maximum light. Fitting a \citet{1989ApJ...345..245C} reddening law the authors also find a preference for  $R_V$ $\sim$ 1.4, significantly smaller than the typical Milky Way value of 3.1. This is confirmed by spectro-polarimetry data in \citet{2015A&A...577A..53P}, who also demonstrate the preference for low $R_V$ in other highly reddened SNe\,Ia, a trend that has been observed with multi-band observations of large samples of nearby SNe \citep{2008A&A...487...19N,2013ApJ...779...38P,2014ApJ...789...32B}.  The low $R_V$ would indicate smaller dust grains in the host of SN~2014J than in the Milky Way. UV spectrophotometry \citep{2015ApJ...805...74B}, the wavelength independence of the polarisation angle \citep{2015A&A...577A..53P} and modeling the optical light curves \citep{2018MNRAS.473.1918B} point towards an interstellar origin of the dust. We note however, that \citet{2014MNRAS.443.2887F} find that a mixture of {\em typical} dust in a combination of interstellar reddening and circumstellar scattering provides good fits to the early multi-wavelength data. For this case an R$_V$ of 2.6 is derived by \citet{2014MNRAS.443.2887F}. We find that the extinction derived from our NIR spectra using the [\ion{Fe}{II}] and [\ion{Co}{II}] line ratios is compatible with maximum light estimates for the reddening. 
 
\section{Conclusions}
\label{sec-conc}

In this study we have presented NIR spectra of SN\,2014J in the nebular phase. The dominant component of these spectra are [\ion{Fe}{II}] and [\ion{Co}{II}] lines. We detect, for the first time, a [\ion{Ni}{II}] line at 1.939 $\mu$m, confirming the presence of stable nickel isotopes. The [\ion{Fe}{II}] and [\ion{Co}{II}] lines are Gaussian with a width of 8\,600\,km\,s$^{-1}$ whereas the [\ion{Ni}{II}] lines at at least as wide and possibly wider at 11\,000\,km\,s$^{-1}$. This indicates that the stable nickel is likely not at low velocities but rather at intermediate velocities, as predicted by multi-D $M_{ch}$ models.  Our line profiles show no evidence for flat-tops. The host galaxy extinction has been estimated from the NIR [\ion{Fe}{II}] line ratios and is seen to be consistent with the inference from near-maximum photometry and polarimetry. Combining our spectral modelling with a prior on the \Nif\,mass from maximum light and $\gamma$-ray observations, we obtain a mass of stable nickel of 0.053\,M$_\odot$.

\begin{acknowledgements}
We would like to thank the staff at Gemini-North, especially Tom Geballe and Rachel Mason for their help during the observations and data reduction. BL and ST acknowledge support by TRR33 "The Dark Universe" of the German Research Foundation. KM acknowledges support from the UK STFC through an Ernest Rutherford Fellowship.

\end{acknowledgements}

\bibliography{Suhail}
\end{document}